\documentclass[twocolumn,english,conference]{IEEEtran}
\usepackage[T1]{fontenc}
\usepackage[latin9]{inputenc}
\usepackage{babel}
\usepackage{amsmath}
\usepackage{amssymb}
\usepackage{graphicx}
\usepackage{setspace}
\usepackage[unicode=true,
 bookmarks=true,bookmarksnumbered=true,bookmarksopen=true,bookmarksopenlevel=1,
 breaklinks=false,pdfborder={0 0 0},backref=false,colorlinks=false]
 {hyperref}
\hypersetup{pdftitle={Your Title},
 pdfauthor={Your Name},
 pdfpagelayout=OneColumn,pdfnewwindow=true,pdfstartview=XYZ,plainpages=false}
\usepackage{breakurl}

\makeatletter
\usepackage{babel}
\@ifundefined{showcaptionsetup}{}{%
 \PassOptionsToPackage{caption=false}{subfig}}
\usepackage{subfig}
\makeatother

\begin{document}

\title{Regularization of the Kernel Matrix via Covariance Matrix Shrinkage
Estimation }

\author{Tomer Lancewicki\\
EBay Inc.\\
625 6th Ave\\
New York, NY\\
Email: tlancewicki@ebay.com}
\maketitle
\begin{abstract}
The kernel trick concept, formulated as an inner product in a feature
space, facilitates powerful extensions to many well-known algorithms.
While the kernel matrix involves inner products in the feature space,
the sample covariance matrix of the data requires outer products.
Therefore, their spectral properties are tightly connected. This allows
us to examine the kernel matrix through the sample covariance matrix
in the feature space and vice versa. The use of kernels often involves
a large number of features, compared to the number of observations.
In this scenario, the sample covariance matrix is not well-conditioned
nor is it necessarily invertible, mandating a solution to the problem
of estimating high-dimensional covariance matrices under small sample
size conditions. We tackle this problem through the use of a shrinkage
estimator that offers a compromise between the sample covariance matrix
and a well-conditioned matrix (also known as the \textquotedbl{}target\textquotedbl{})
with the aim of minimizing the mean-squared error (MSE). We propose
a distribution-free kernel matrix regularization approach that is
tuned directly from the kernel matrix, avoiding the need to address
the feature space explicitly. Numerical simulations demonstrate that
the proposed regularization is effective in classification tasks. 
\end{abstract}

\section{Introduction}

Machine learning tasks often require a data pre-processing stage also
known as feature extraction. One common and powerful way of extracting
features is by employing the \textquotedbl{}kernel trick\textquotedbl{}
concept, formulated as an inner product in a feature space, which
allows us to build interesting extensions to existing algorithms.
The general idea is that, if an algorithm can be formulated in a way
that the input vector enters only in a scalar product form, then that
scalar product can be replaced with some other choice of kernel. Many
models for regression and classification can be reformulated in terms
of a dual representation in which the kernel function arises naturally.
For instance, the kernel trick technique can be applied to principal
component analysis in order to develop a nonlinear variant of \emph{principal
component analysis} (PCA) \cite{scholkopf1998nonlinear,Lancewicki2014382},
while other examples include \emph{nearest-neighbor} classifiers \cite{nearestneighbor2014},
\emph{kernel Fisher discriminant analysis} (KFDA)\cite{scholkopft1999fisher}
and \emph{support vector machines} (SVM's) \cite{scholkopf2002learning}.

The aforementioned schemes attempt to discover structure in the data.
For example, in pattern recognition and regression estimation, we
are given a training set of inputs and outputs, and attempt to infer
the test outputs for unseen inputs. This is only possible if we have
some measure for determining how the test set is related to the train
set. Generally speaking, we expect similar inputs to lead to similar
outputs whereby similarity is commonly measured in terms of a loss
function. The latter indicates how well the inferred outputs match
the true outputs. The training stage commonly involves a risk function
that contains a term measuring the loss incurred for the training
patterns. However, in order to generalize well to the test data, it
is also necessary to control the complexity of the model used for
explaining the training data, a task that is often accomplished with
the help of regularization terms \cite[Ch. 4]{scholkopf2002learning}.
Minimizing the empirical risk without regularization terms can lead
to numerical instabilities and poor generalization performance. Therefore,
it is essential to utilize objective functions that involve both the
empirical loss term and a regularization term. A possible way to avoid
the aforementioned problems is to use the class of admissible solutions
\cite{tikhonov2013numerical}, often referred to in statistics as
shrinkage estimators \cite{james1961estimation}.

In this paper we point out the connections between regularization
of the kernel matrix and a shrinkage estimation of the covariance
matrix in the feature space. Since the kernel matrix and the sample
covariance matrix involve inner and outer products in the feature
space, respectively, their spectral properties are tightly connected
\cite{tradeoff2013}. This allows us to examine the kernel matrix
stability through the sample covariance matrix in the feature space
and vice versa. More specifically, the use of kernels often involve
a large number of features, compared to the number of observations.
In this scenario, the sample covariance matrix is not well-conditioned
nor is it necessarily invertible (despite the fact that those two
properties are required for most machine learning applications). This
necessitates that a solution be found for the problem of estimating
high-dimensional covariance matrices under small sample size settings.

There already exists an extensive body of literature concerning the
small sample size scenario \cite{bickel,rohde2011,lancewicki_RL,lancewicki_control,lancewicki2015sequential},
achieved by incorporating additional knowledge into the estimation
process, such as sparseness \cite{ravikumar2011,icml2014_2,cai2012minimax},
a graph model \cite{rajaratnam2008}, \cite{khare2011}, a factor
model \cite{Fan2008186}, or other references therein. However, such
additional knowledge is often either unavailable or is not trustworthy.
In the absence of further knowledge about the structure of the true
covariance matrix, the most successful approach so far has been shrinkage
estimation \cite{ledoit2011nonlinear}.

\begin{singlespace}
This paper proposes an analytic distribution-free regularization of
the kernel matrix through a shrinkage estimation of the sample covariance
matrix, which is optimal in the sense of mean-squared error (MSE).
The regularization can be utilized directly from the kernel matrix,
therefore releasing us from dealing with the feature space explicitly.

The rest of the paper is organized as follows. In Section 2, we formulate
the problem. In Section 3 we introduce the shrinkage estimation and
derive its optimal solution with respect to the kernel matrix. In
Section 4 we examine the relation between the kernel matrix and the
sample covariance matrix. Section 5 presents numerical simulation
results for classification tasks. Section 6 summarizes our principal
conclusions. To make for easier reading, the derivations of some of
our results appear in the appendix. 
\end{singlespace}

Notation: We depict vectors in lowercase boldface letters and matrices
in uppercase boldface. The transpose operator is denoted as $\left(\cdot\right)^{T}$.
The trace and the Frobenius norm of a matrix are denoted as $\mathrm{Tr}\left(\cdot\right)$
and $\left\Vert \cdot\right\Vert _{F}$, respectively. The identity
matrix is denoted as $\mathbf{I}$, while $\mathbf{e}=\left[1,1,\ldots,1\right]^{T}$
is a column vector of all ones, and $\mathbf{1}=\mathbf{e}\mathbf{e}^{T}$
is a matrix of ones. The centering matrix is denoted as $\mathbf{H}=\mathbf{I}-\mathbf{1}/n$.
For any real matrices $\mathbf{R}_{1}$ and $\mathbf{R}_{2}$, the
inner product is defined as $\left\langle \mathbf{R}_{1},\mathbf{R}_{2}\right\rangle =\mathrm{Tr}\left(\mathbf{R}_{1}^{T}\mathbf{R}_{2}\right)$,
where $\left\langle \mathbf{R}_{1},\mathbf{R}_{1}\right\rangle =\left\Vert \mathbf{R}_{1}\right\Vert _{F}^{2}$
\cite[Sec. 2.20]{MatrixHandbook}. To make for easier reading, when
$\mathbf{R}_{1}$ is a random matrix, we use the notation $V(\mathbf{R}_{1})=E\left\{ \left\Vert \mathbf{R}_{1}-E\left\{ \mathbf{R}_{1}\right\} \right\Vert _{F}^{2}\right\} $
(the sum of variances of the elements in $\mathbf{R}_{1}$), where
$E\left\{ \cdot\right\} $ denotes the expectation operator.

\section{Problem Formulation}

Let $\left\{ \mathbf{x}_{i}\right\} _{i=1}^{n},\,\mathbf{x}_{i}\in\mathbb{R}^{q}$
be a sample of \emph{independent identical distributed} (i.i.d.) $q$-dimensional
vectors, and let $\boldsymbol{\phi}\left(\mathbf{x}\right)$ be a
non-linear mapping to a $p$-dimensional feature space with a covariance
matrix $\boldsymbol{\Sigma}$ of size $p\times p$. When the number
of observations $n$ is large (i.e., $n\gg p$), the most common estimator
of $\boldsymbol{\Sigma}$ is the sample covariance matrix $\mathbf{S}$
of size $p\times p$, defined as 
\begin{equation}
\mathbf{S}=\frac{1}{n-1}\boldsymbol{\Phi}\mathbf{H}\boldsymbol{\Phi}^{T},\label{eq:Smatrix-1}
\end{equation}
where $\boldsymbol{\Phi}$ is the $p\times n$ matrix design matrix
whose $i^{th}$ column is given by $\boldsymbol{\phi}_{i}=\boldsymbol{\phi}\left(\mathbf{x}_{i}\right)$.
The matrix $\mathbf{S}$ \eqref{eq:Smatrix-1} is an unbiased estimator
of $\boldsymbol{\Sigma}$ , i.e., $E\left\{ \mathbf{S}\right\} =\boldsymbol{\Sigma}$.

The kernel function in the feature space $\boldsymbol{\phi}\left(\mathbf{x}\right)$
is given by the relation 
\begin{equation}
k\left(\mathbf{x}_{i},\mathbf{x}_{j}\right)=\boldsymbol{\phi}\left(\mathbf{x}_{i}\right)^{T}\boldsymbol{\phi}\left(\mathbf{x}_{j}\right),
\end{equation}
where the kernel matrix of size $n\times n$ is defined by 
\begin{equation}
\mathbf{K}=\mathbf{H}\boldsymbol{\Phi}^{T}\boldsymbol{\Phi}\mathbf{H}.\label{eq:Smatrix-1-1}
\end{equation}
Multiplication by the centering matrix $\mathbf{H}$ makes the data
centered in the feature space \cite[Ch. 14.2]{scholkopf2002learning}.
Since the use of kernels mostly involves large number of features
$p$, compared to the number of observations $n$, we are forced to
deal with the problem of estimating high dimensional covariance matrices
under a small sample size. In this scenario, the sample covariance
matrix $\mathbf{S}$ \eqref{eq:Smatrix-1} is not well-conditioned
nor is it necessarily invertible. When $n\leq p$, the inversion cannot
be computed at all \cite[Sec. 2.2]{RLDM1}. Although significant progress
in dealing with the small sample size problem has been made by various
regularization methods \cite{bickel,slda1,SMT}, for example, by improving
the accuracy of the pseudo-inverse of $\mathbf{S}$ \cite{Hoyle,Raudys}
or by regularizing $\mathbf{S}$ for specific tasks in discriminant
analysis \cite{slda2,kernelQDA,optimizationCriterion1,OptimizationCriterion2,slda7};
these methods require the explicit expression of the feature space
$\boldsymbol{\phi}\left(\mathbf{x}\right)$. In general, when using
kernels, the feature space is only known implicitly.

In the following section, we develop an analytical solution to regularized
the kernel matrix $\mathbf{K}$ \eqref{eq:Smatrix-1-1} by examining
the relationship between $\boldsymbol{\Sigma}$, $\mathbf{S}$ \eqref{eq:Smatrix-1}
and $\mathbf{K}$ \eqref{eq:Smatrix-1-1} through a shrinkage estimation
for covariance matrices. It has been demonstrated by \cite{slda8}
that the largest sample eigenvalues of $\mathbf{S}$ are systematically
biased upwards, while the smallest ones downwards. This bias can be
corrected by pulling down the largest eigenvalues and pushing up the
smallest ones, toward the grand mean of all sample eigenvalues. We
tackle this problem through the use of a shrinkage estimator that
offers a compromise between the sample covariance matrix and a well-conditioned
matrix (also known as the \textquotedbl{}target\textquotedbl{}) with
the aim of minimizing the mean-squared error (MSE). Since the spectral
properties of $\mathbf{S}$ \eqref{eq:Smatrix-1} and $\mathbf{K}$
\eqref{eq:Smatrix-1-1} are tightly connected, any modification of
the eigenvalues in $\mathbf{S}$ automatically result a related change
on the eigenvalues of $\mathbf{K}$ that reflect the optimal estimation
of the covariance matrix in the feature space, in the sense of MSE.

\section{Shrinkage Estimator for Covariance Matrices}

We briefly review the topic of single-target shrinkage estimator for
an unknown covariance matrix $\boldsymbol{\Sigma}$ by following \cite{slda8,slda12},
which is generally applied to high-dimensional estimation problems.
The shrinkage estimator $\hat{\mathbf{\boldsymbol{\Sigma}}}\left(\lambda\right)$
is in the form 
\begin{equation}
\hat{\mathbf{\boldsymbol{\Sigma}}}\left(\lambda\right)=(1-\lambda)\mathbf{S}+\lambda\mathbf{T}\label{eq:estimator}
\end{equation}
where the target $\mathbf{T}$ is a restricted estimator of $\boldsymbol{\Sigma}$
defined as 
\begin{equation}
\mathbf{T}=\frac{\mathrm{Tr}\left(\mathbf{S}\right)}{p}\mathbf{I}.\label{eq:Tar1}
\end{equation}
The objective is to find an estimator $\hat{\mathbf{\boldsymbol{\Sigma}}}\left(\lambda\right)$
which minimizes the \emph{mean squared error} (MSE) 
\begin{equation}
E\left\{ \left\Vert \hat{\mathbf{\boldsymbol{\Sigma}}}\left(\lambda\right)-\boldsymbol{\Sigma}\right\Vert _{F}^{2}\right\} .\label{eq:MSE}
\end{equation}
The value of $\lambda$ that minimize the MSE \eqref{eq:MSE} is defined
as 
\begin{equation}
\lambda_{O}=\arg\min_{\lambda}E\left\{ \left\Vert \hat{\mathbf{\boldsymbol{\Sigma}}}\left(\lambda\right)-\boldsymbol{\Sigma}\right\Vert _{F}^{2}\right\} \label{eq:lambdaO-3}
\end{equation}
and can be given by the distribution-free formula 
\begin{equation}
\lambda_{O}=\frac{E\left\{ \left\langle \mathbf{T}-\mathbf{S},\boldsymbol{\Sigma}-\mathbf{S}\right\rangle \right\} }{E\left\{ \left\Vert \mathbf{T}-\mathbf{S}\right\Vert _{F}^{2}\right\} }.\label{eq:lambdaO-1}
\end{equation}
The scalar $\lambda_{O}$ is called the oracle shrinkage coefficient,
since it depends on the unknown covariance matrix $\boldsymbol{\Sigma}$.
Therefore, $\lambda_{O}$ \eqref{eq:lambdaO-1} must be estimated.
As we will show next, the optimal shrinkage coefficient of the covariance
matrix can be estimated directly from the kernel matrix $\mathbf{K}$
\eqref{eq:Smatrix-1-1}, without explicitly dealing with $\boldsymbol{\phi}\left(\mathbf{x}\right)$,
commonly unknown in practice.

\subsection{{\normalsize{}{}Estimations of the Oracle Shrinkage Coefficient
}$\lambda_{O}$ \eqref{eq:lambdaO-1} {\normalsize{}{}\label{sec:The-sample-counterpart}}}

In this section, we show that the unbiased estimator of $\lambda_{O}$
\eqref{eq:lambdaO-1-1}, denoted as $\hat{\lambda}_{O}$, can be written
as a function of the kernel matrix $\mathbf{K}$ \eqref{eq:Smatrix-1-1}.
It has been shown in \cite[Sec. 3.B]{Lance} that the target $\mathbf{T}$
\eqref{eq:Tar1} is a private case of the general target framework
which allows to reformulate $\lambda_{O}$ \eqref{eq:lambdaO-1} as
\begin{equation}
\lambda_{O}=\frac{V(\mathbf{S})-V(\mathbf{T})}{E\left\{ \left\Vert \mathbf{T}-\mathbf{S}\right\Vert _{F}^{2}\right\} }.\label{eq:lambdaO-1-1}
\end{equation}
The oracle shrinkage coefficient $\lambda_{O}$ \eqref{eq:lambdaO-1-1}
can be estimated from its sample counterparts as 
\begin{equation}
\hat{\lambda}_{O}=\max\left(\min\left(\frac{\hat{V}(\mathbf{S})-\hat{V}(\mathbf{T})}{\left\Vert \mathbf{T}-\mathbf{S}\right\Vert _{F}^{2}},1\right),0\right),\label{eq:lambdaO_est-1}
\end{equation}
where the symbol \textasciicircum{} indicates an estimated value of
the parameter. The estimated oracle shrinkage coefficient $\hat{\lambda}_{O}$
\eqref{eq:lambdaO_est-1} is bounded in {[}0,1{]} in order to keep
the shrinkage estimator $\hat{\mathbf{\boldsymbol{\Sigma}}}\left(\hat{\lambda}_{O}\right)$
positive-definite as required from a covariance matrix \cite{Lance,lancewicki2014dimensionality}.
The unbiased estimators of $\hat{V}(\mathbf{S})$ and $\hat{V}(\mathbf{T})$
are derived in appendix A and can be written as a function of the
kernel matrix $\mathbf{K}$ \eqref{eq:Smatrix-1-1}, i.e., 
\begin{equation}
\hat{V}(\mathbf{S})=\frac{n}{\left(n-1\right)^{2}\left(n-2\right)}\left(\left\Vert diag\left(\mathbf{K}\right)\right\Vert _{F}^{2}-\frac{1}{n}\left\Vert \mathbf{K}\right\Vert _{F}^{2}\right)\label{eq:vs-1}
\end{equation}
and 
\begin{equation}
\hat{V}(\mathbf{T})=\frac{n}{p\left(n-1\right)^{2}\left(n-2\right)}\left\Vert diag\left(\mathbf{K}\right)-\frac{1}{n}\mathrm{Tr}\left(\mathbf{K}\right)\mathbf{e}\right\Vert _{F}^{2},\label{eq:vt2}
\end{equation}
respectively. The denominator of $\hat{\lambda}_{O}$ \eqref{eq:lambdaO_est-1}
can be also written as a function of $\mathbf{K}$ \eqref{eq:Smatrix-1-1},
i.e., 
\begin{equation}
\left\Vert \mathbf{T}-\mathbf{S}\right\Vert _{F}^{2}=\frac{1}{\left(n-1\right)^{2}}\left(\left\Vert \mathbf{K}\right\Vert _{F}^{2}-\frac{1}{p}\mathrm{Tr}^{2}\left(\mathbf{K}\right)\right).\label{eq:ts12}
\end{equation}
Therefore, by using $\hat{V}(\mathbf{S})$ \eqref{eq:vs-1}, $\hat{V}(\mathbf{T})$
\eqref{eq:vt2} and \eqref{eq:ts12}, the estimated oracle shrinkage
coefficient $\hat{\lambda}_{O}$ \eqref{eq:lambdaO_est-1} can be
written as a function of the kernel matrix $\mathbf{K}$ \eqref{eq:Smatrix-1-1};
importantly, without dealing explicitly with the feature space $\boldsymbol{\phi}\left(\mathbf{x}\right)$.
In the next section we utilize the shrinkage coefficient $\hat{\lambda}_{O}$
\eqref{eq:lambdaO_est-1} in order to regularized the kernel matrix
$\mathbf{K}$ \eqref{eq:Smatrix-1-1}.

\section{Kernel regularization through shrinkage estimation}

In this section we examine the relation between $\boldsymbol{\Sigma}$,
$\mathbf{S}$ \eqref{eq:Smatrix-1} and $\mathbf{K}$ \eqref{eq:Smatrix-1-1}
with respect to the shrinkage estimator $\hat{\mathbf{\boldsymbol{\Sigma}}}\left(\lambda\right)$
\eqref{eq:estimator}. Let denote $\zeta_{i},i=1,\ldots,p$ as the
eigenvalues of the unknown covariance matrix $\boldsymbol{\Sigma}$
in decreasing order, i.e., $\zeta_{1}\geq\zeta_{2}\geq\ldots\geq\zeta_{p}$.
It is well known that $\sum_{i=1}^{p}\zeta_{i}=\mathrm{Tr}\left(\boldsymbol{\Sigma}\right)$
\cite[Ch. 6.17]{MatrixHandbook}. As a result, the squared bias of
$\mathbf{T}$ \eqref{eq:Tar1} with respect to $\boldsymbol{\Sigma}$
can be written as 
\begin{equation}
\left\Vert E\left\{ \mathbf{T}\right\} -\boldsymbol{\Sigma}\right\Vert _{F}^{2}=\left\Vert \frac{1}{p}\mathrm{Tr}\left(\boldsymbol{\Sigma}\right)\mathbf{I}-\mathbf{V}\boldsymbol{\Lambda}\mathbf{V}^{T}\right\Vert _{F}^{2}=\sum_{i=1}^{p}\left(\zeta_{i}-\bar{\zeta}\right)^{2}\label{eq:b1}
\end{equation}
where $\bar{\zeta}$ is the mean of the eigenvalues $\zeta_{i},i=1,\ldots,p$,
i.e., 
\begin{equation}
\bar{\zeta}=\frac{\mathrm{Tr}\left(\boldsymbol{\Sigma}\right)}{p}=\frac{1}{p}\sum_{i=1}^{p}\zeta_{i},\label{eq:eigmean}
\end{equation}
and the matrices $\mathbf{V},\boldsymbol{\Lambda}$ are the eigenvector
and eigenvalue matrices of $\boldsymbol{\Sigma}$, respectively, such
that $\boldsymbol{\Sigma}=\mathbf{V}\boldsymbol{\Lambda}\mathbf{V}^{T}$.

The above result shows that the squared bias of $\mathbf{T}$ \eqref{eq:Tar1},
i.e., $\left\Vert E\left\{ \mathbf{T}\right\} -\boldsymbol{\Sigma}\right\Vert _{F}^{2}$
\eqref{eq:b1}, is equal to the dispersion of the eigenvalues around
their mean. Let denote $\delta_{i},i=1,\ldots,p$ as the eigenvalues
of $\mathbf{S}$ \eqref{eq:Smatrix-1} in decreasing order, i.e.,
$\delta_{1}\geq\delta_{2}\geq\ldots\geq\delta_{p}$. Using \eqref{eq:b1},
the eigenvalues dispersion of $\mathbf{S}$ around their mean is equal
to 
\begin{equation}
E\left\{ \left\Vert \mathbf{S}-\bar{\zeta}\mathbf{I}\right\Vert _{F}^{2}\right\} =E\left\{ \sum_{i=1}^{p}\left(\delta_{i}-\bar{\zeta}\right)^{2}\right\} =V(\mathbf{S})+\sum_{i=1}^{p}\left(\zeta_{i}-\bar{\zeta}\right)^{2},
\end{equation}
indicate that the eigenvalues of $\mathbf{S}$ are more dispersed
around their mean then the true ones, where the excess dispersion
is equal to $V(\mathbf{S})$. The excess dispersion implies that the
largest eigenvalues of $\mathbf{S}$ are biased upwards while the
smallest downwards. Therefore, we can improve upon the sample covariance
matrix by shrinking its eigenvalues toward their mean $\bar{\zeta}$
\eqref{eq:eigmean}. This is done practically via the shrinkage estimator
$\hat{\mathbf{\boldsymbol{\Sigma}}}\left(\lambda\right)$ \eqref{eq:estimator}
where the optimal shrinkage coefficient in the sense of MSE is equal
to $\hat{\lambda}_{O}$ \eqref{eq:lambdaO_est-1}.

The above results relates to the kernel matrix $\mathbf{K}$ \eqref{eq:Smatrix-1-1}
as follows. Let denote $\kappa_{i},i=1,\ldots,n$ as the eigenvalues
of $\mathbf{K}$ \eqref{eq:Smatrix-1-1} in decreasing order, i.e.,
$\kappa_{1}\geq\kappa_{2}\geq\ldots\geq\kappa_{n}$. In the small
sample size scenario ($n<p$) , it is straight forward to show that
the eigenvalues of $\mathbf{K}$ \eqref{eq:Smatrix-1-1} and $\mathbf{S}$
\eqref{eq:Smatrix-1} are related by 
\begin{equation}
\kappa_{i}=\left(n-1\right)\delta_{i},\, i=1,\ldots,n-1.\label{eq:relation}
\end{equation}
The other $\left(p-n+1\right)$ eigenvalues of $\mathbf{S}$ are all
zero and their eigenvectors are indefinite.

The procedure of regularize the kernel matrix $\mathbf{K}$ \eqref{eq:Smatrix-1-1}
by 
\begin{equation}
\hat{\mathbf{\mathbf{K}}}\left(\hat{\lambda}_{O}\right)=(1-\hat{\lambda}_{O})\mathbf{K}+\hat{\lambda}_{O}\frac{\mathrm{Tr}\left(\mathbf{K}\right)}{p}\mathbf{I}\label{eq:estimator-1}
\end{equation}
is therefore equivalent to the correction of the first $n$ eigenvalues
of the sample covariance matrix $\mathbf{S}$ \eqref{eq:Smatrix-1}
with respect to their mean in the feature space, which is optimal
in the sense of MSE. We examine the regularized kernel matrix $\hat{\mathbf{\mathbf{K}}}\left(\hat{\lambda}_{O}\right)$
\eqref{eq:estimator-1} in the next section.

\section{Experiments}

To provide some insight into how the regularized kernel matrix $\hat{\mathbf{\mathbf{K}}}\left(\hat{\lambda}_{O}\right)$
\eqref{eq:estimator-1} behaves, we consider a two-class classification
problem. The first class observations are generated from a two Gaussians
in a two-dimensional space with standard deviation of 0.1, centered
at $\left(-0.5,-0.2\right)$ and $\left(0.5,0\right)$. The second
class observations are generated from a two-dimensional Gaussian with
standard deviation of 0.1, centered at $\left(0,0\right)$. The problem
is not linearly separable, giving raise to the use of kernels. In
order to distinguish between the two classes, the \emph{kernel Fisher
discriminant analysis} (KFDA) \cite{scholkopft1999fisher} is used
by utilizing the radial basis function kernel 
\begin{equation}
k\left(\mathbf{x}_{i},\mathbf{x}_{j}\right)=\exp\left(-\frac{\left\Vert \mathbf{x}_{i}-\mathbf{x}_{j}\right\Vert _{F}^{2}}{2\sigma^{2}}\right)\label{eq:kernel}
\end{equation}
with $\sigma^{2}=0.1$. We regularize the within-class scatter of
the KFDA using $\hat{\mathbf{\mathbf{K}}}\left(\hat{\lambda}_{O}\right)$
\eqref{eq:estimator-1} versus the proposed fixed regularization of
$\lambda=10^{-3}$ \cite{scholkopft1999fisher}. These two scenarios
are referred to as ``shrinkage KFDA'' and ``KFDA'', respectively.
We run the experiments when the number of training observation per
Gaussian, denoted by $n_{g}$, varies from 3 to 30. Each simulation
is repeated 100 times and the average values of the misclassification
rates as a function of $n_{g}$ are depicted in Fig. 1.

\begin{figure}[tbh]
\begin{centering}
\textsf{\includegraphics[scale=0.67]{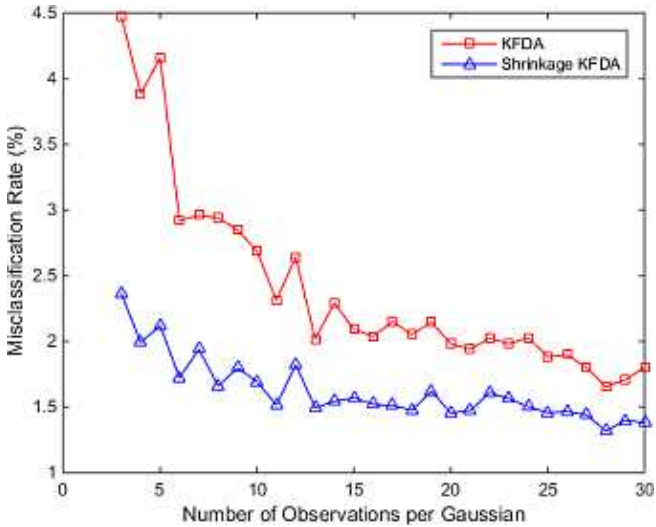}} 
\par\end{centering}

\protect\protect\caption{Misclassification rates of KFDA versus shrinkage KFDA as a function
of the number of training observations per Gaussian }
\end{figure}

As can be seen from Fig. 1, the shrinkage KFDA outperforms the KFDA
for any value of $n_{g}$. We used the paired t-test in order to evaluate
the null hypothesis in which the difference between the two misclassification
rates over the 100 simulations comes from a population with mean equal
to zero. The t-test rejects the null hypothesis at the 99\% significance
level, meaning that the shrinkage KFDA consistently outperformed the
KFDA.

The average value of $\hat{\lambda}_{O}$ \eqref{eq:lambdaO_est-1}
as a function of $n_{g}$ (training observations per Gaussian) is
plotted in Fig. 2. When considering a small number of observations,
the shrinkage coefficient is relatively high, giving rise to the well
structured target $\mathbf{T}$ \eqref{eq:Tar1}. As the number of
observations increases, using $\mathbf{S}$ \eqref{eq:Smatrix-1}
is preferred, primarily since it provides a more accurate description
of the true covariance matrix $\boldsymbol{\Sigma}$. Consequently,
the shrinkage coefficient $\hat{\lambda}_{O}$ \eqref{eq:lambdaO_est-1}
decreases as the number of training observations $n_{g}$ increase.

\begin{figure}[tbh]
\begin{centering}
\textsf{\includegraphics[scale=0.67]{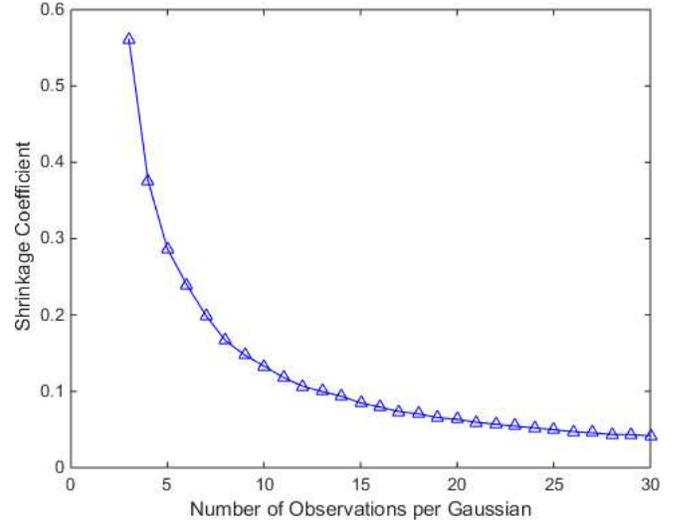}} 
\par\end{centering}

\protect\protect\caption{Shrinkage coefficient $\hat{\lambda}_{O}$ \eqref{eq:lambdaO_est-1}
as a function of the number of training observations per Gaussian }
\end{figure}

In order to visually examine the impact of regularization, we present
the KFDA decision boundaries where the number of observations for
each Gaussian $n_{g}$ is 5. In other words, the first class contains
10 training observations while the second group includes 5 training
observations. Fig. 3 depicts the classification performance resulting
from training the KFDA and shrinkage KFDA on the data set using the
kernel \eqref{eq:kernel}. Fig. 3(a) shows the output value produced
by the KFDA from inputs in the two-dimensional grid. The outputs produce
contour lines of constant values that varies monotonically along the
underlying nonlinear structure of the data. Fig. 3(b) illustrates
the decision boundaries found by KFDA. It can be seen clearly that
although the data set is not linearly separable in the two-dimensional
data space, it is linearly separable in the nonlinear feature space
defined implicitly by the nonlinear kernel \eqref{eq:kernel}. Hence,
the training data points are perfectly separated in the original data
space. However, it can be seen from Fig. 3(b) that the decision boundaries
stretched far beyond the area surrounding the second class. This is
the result of a poor regularization term used by the KFDA, leading
to over fitting with respect to the training observations and bad
generalization performance. In comparison, Fig. 3(c) shows the output
value produced by the shrinkage KFDA from inputs in the two-dimensional
grid. Again, the outputs produces contour lines of constant values
that varies monotonically along the underlying nonlinear structure
of the data. The contour lines in Fig. 3(c) follows the data structure
more moderately than the contour lines in Fig. 3(a). Fig. 3(d) present
the decision boundaries found by shrinkage KFDA. Again, the data set
is linearly separable in the nonlinear feature space. The decision
boundaries in Fig. 3(d) does not over fit to the training data as
in Fig. 3(b).

The same experiment is repeated where the number of observations for
each Gaussian $n_{g}$ is equal 20, i.e., the first class have 40
training observations while the second group have 20 training observations.
The results are provided in Fig. 4. Although the training data points
are perfectly separated in the original data space both by KFDA and
by shrinkage KFDA; the KFDA overfits the training data. As a consequence,
the decision boundaries produces by the KFDA creates three different
areas that relates to class 2. This is again the result of a poor
regularization term. The shrinkage KFDA contour lines in Fig. 4(c)
follow the data more moderately than in Fig. 4(a). As a result, the
decision boundaries found by the shrinkage KFDA, shown in Fig. 4(d),
clearly separate the two classes, without overfitting to the training
data as in Fig. 4(b).

\begin{figure}[tbh]
\begin{centering}
\subfloat[]{\includegraphics[scale=0.36]{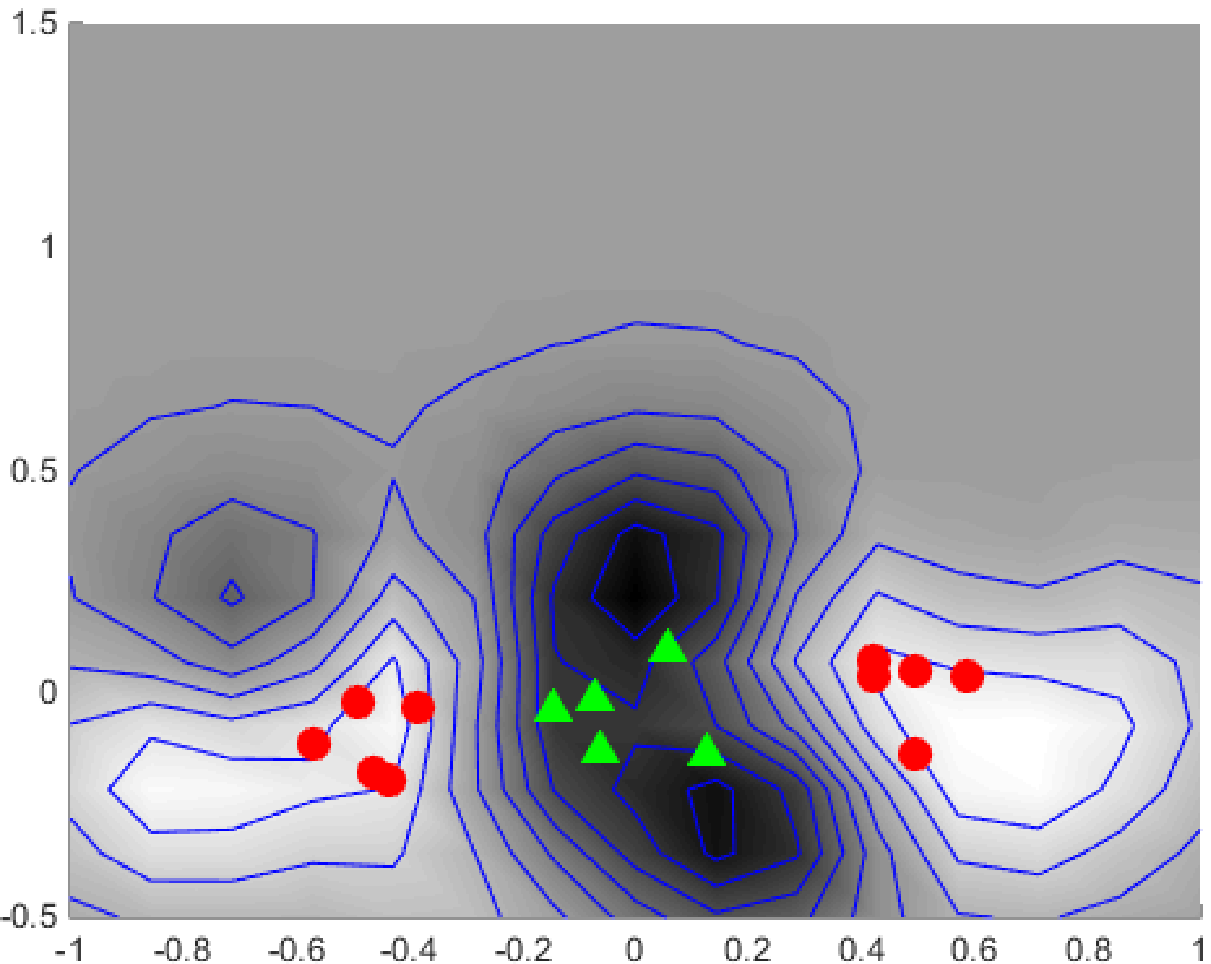}

}\subfloat[]{\includegraphics[scale=0.36]{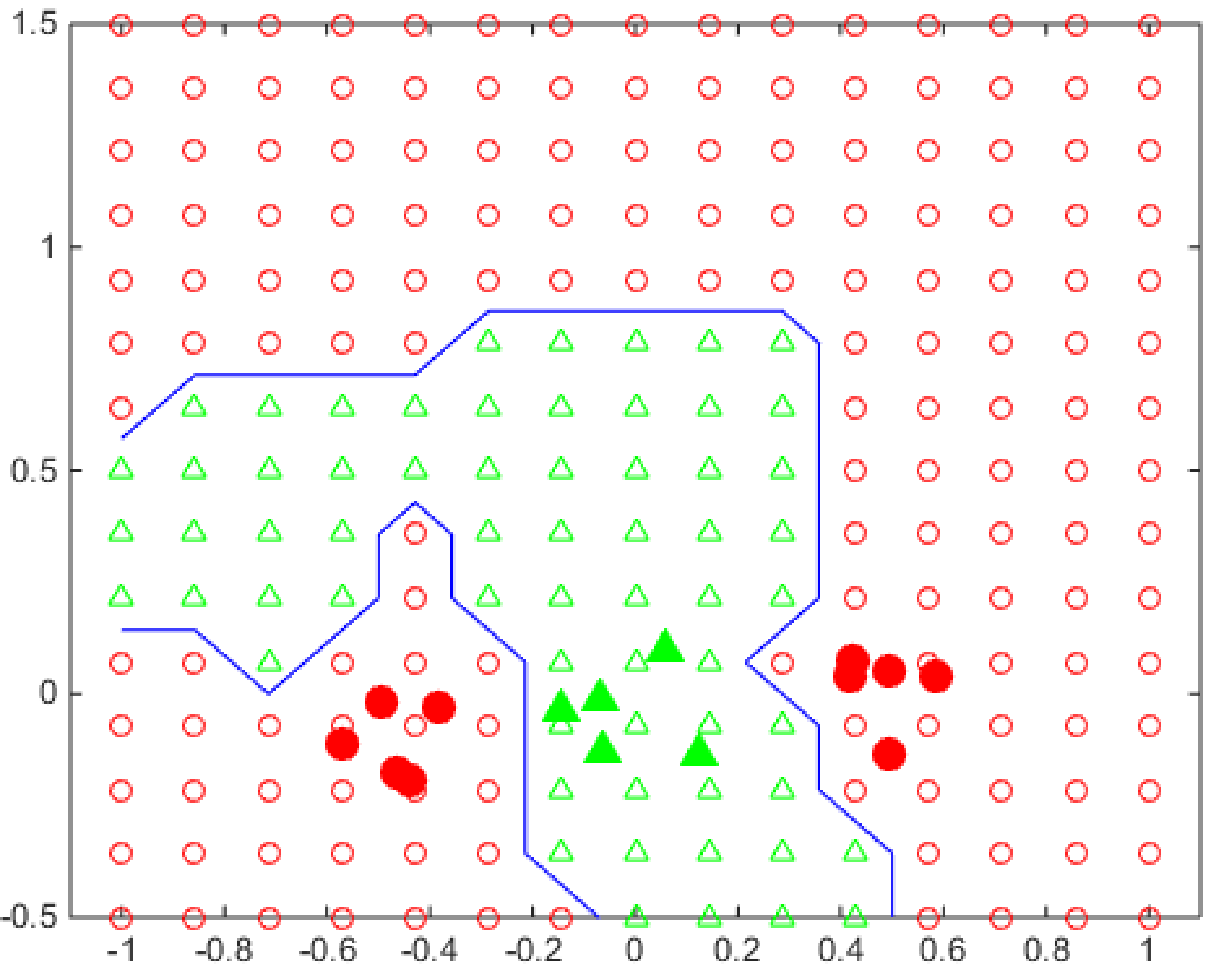}

}
\par\end{centering}

\begin{centering}
\subfloat[]{\includegraphics[scale=0.36]{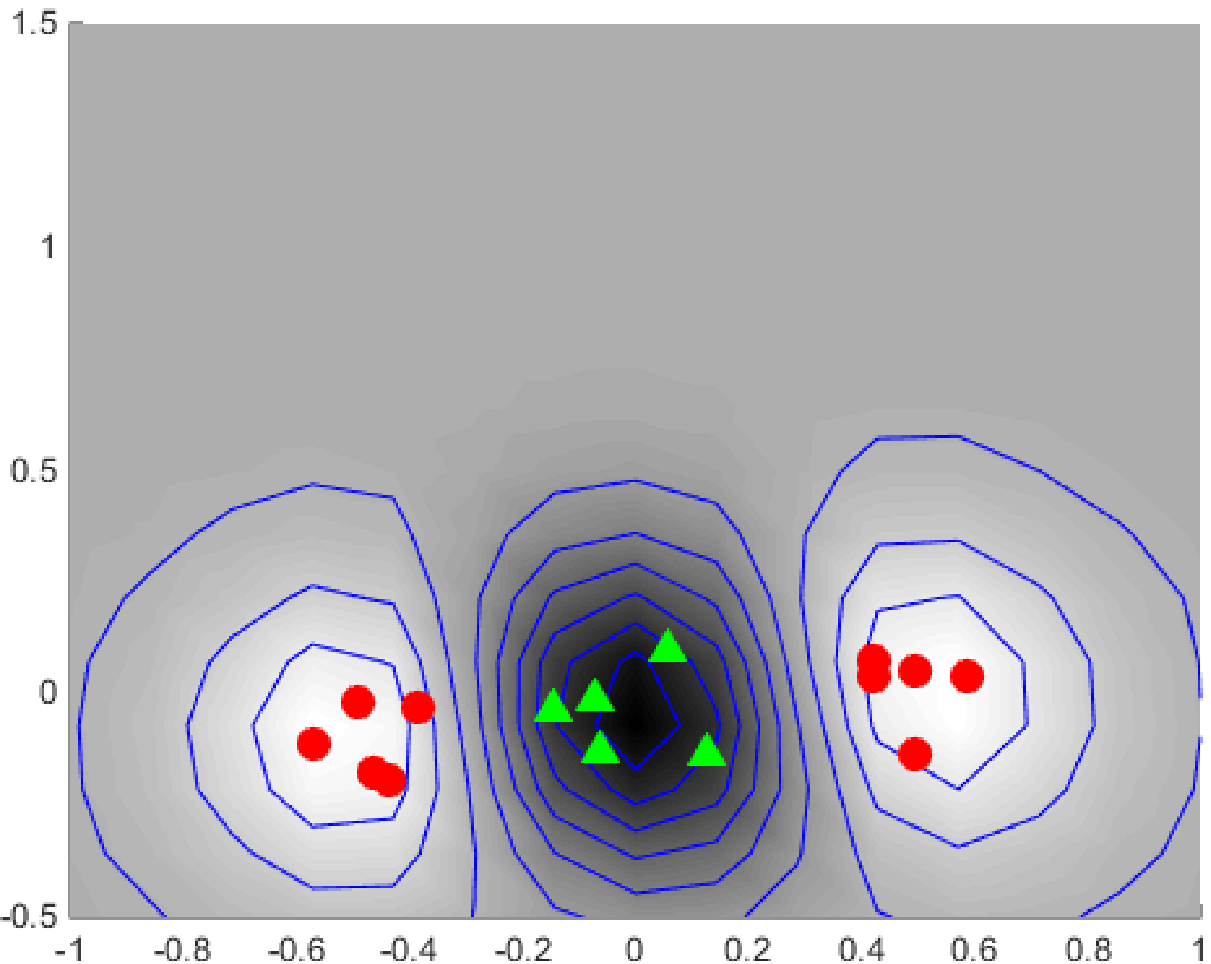}

}\subfloat[]{\includegraphics[scale=0.36]{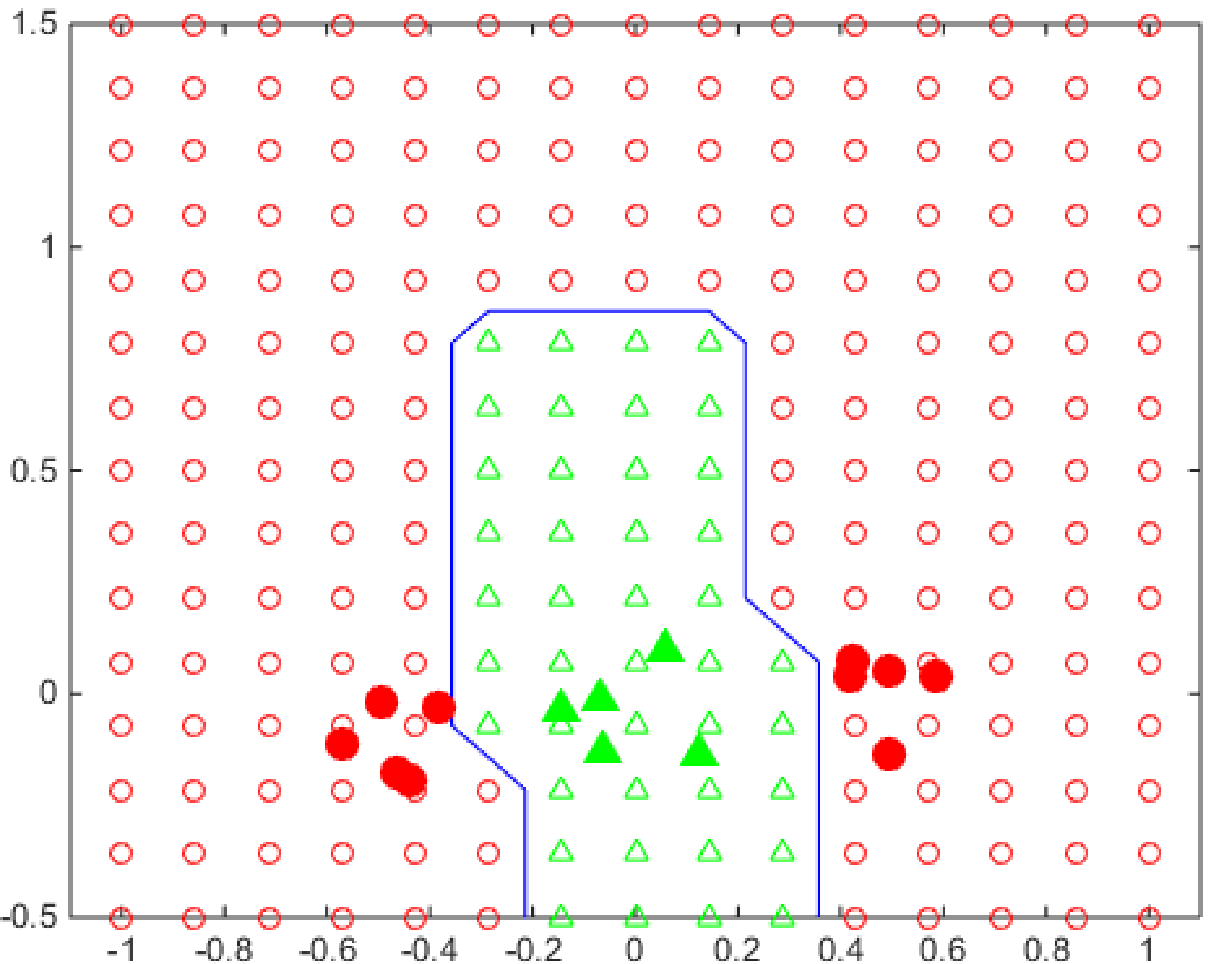}

}
\par\end{centering}

\protect\protect\caption{Classification results for KFDA and Shrinkage KFDA using the kernel
\eqref{eq:kernel} where the number of observations for each Gaussian
$n_{g}$ is equal 5. (a) Contours of constant outputs obtained from
KFDA. (b) Decision boundaries obtained from KFDA. (c) Contours of
constant outputs obtained from shrinkage KFDA. (d) Decision boundaries
obtained from shrinkage KFDA. }
\end{figure}

\begin{figure}[tbh]
\begin{centering}
\subfloat[]{\includegraphics[scale=0.36]{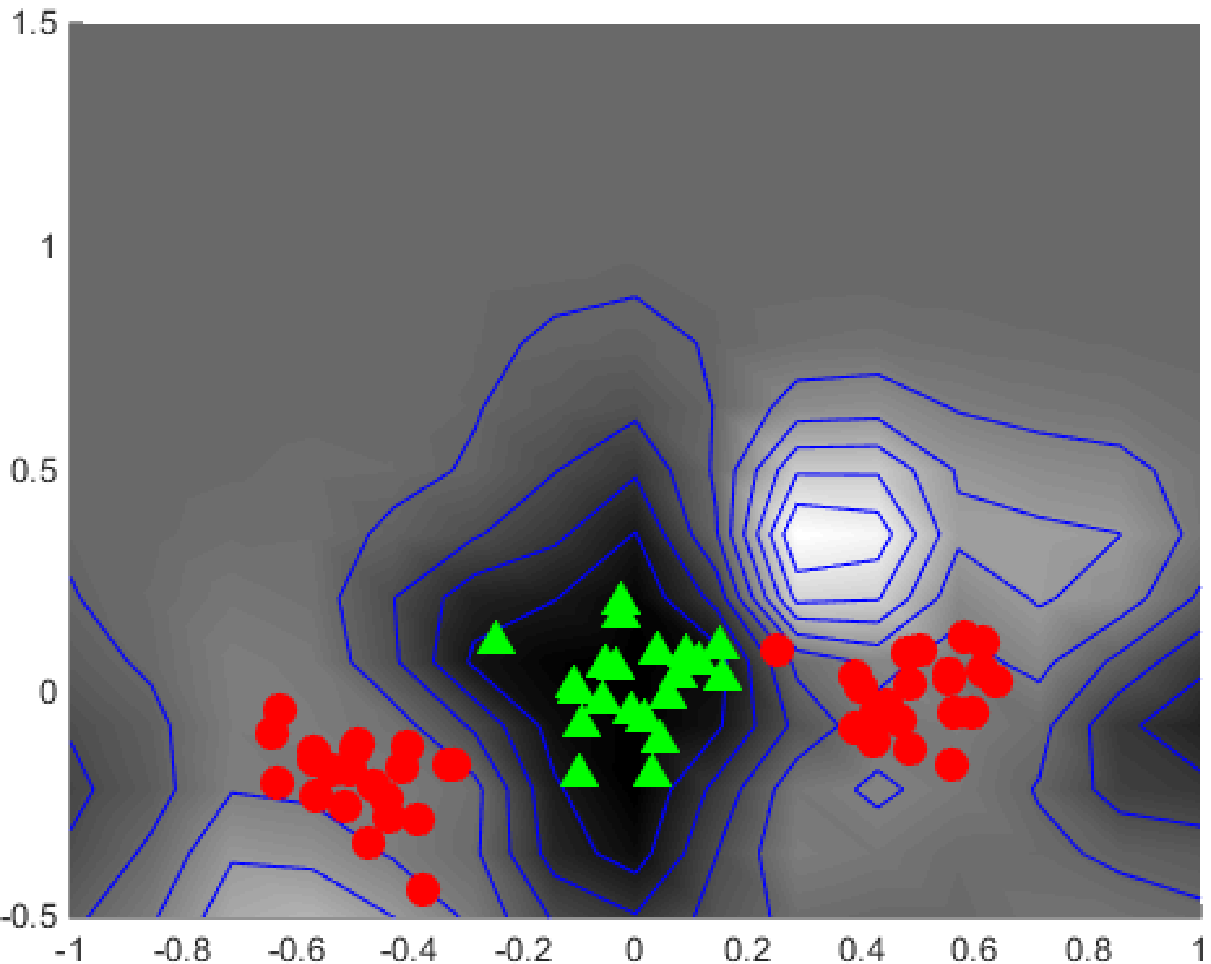}

}\subfloat[]{\includegraphics[scale=0.36]{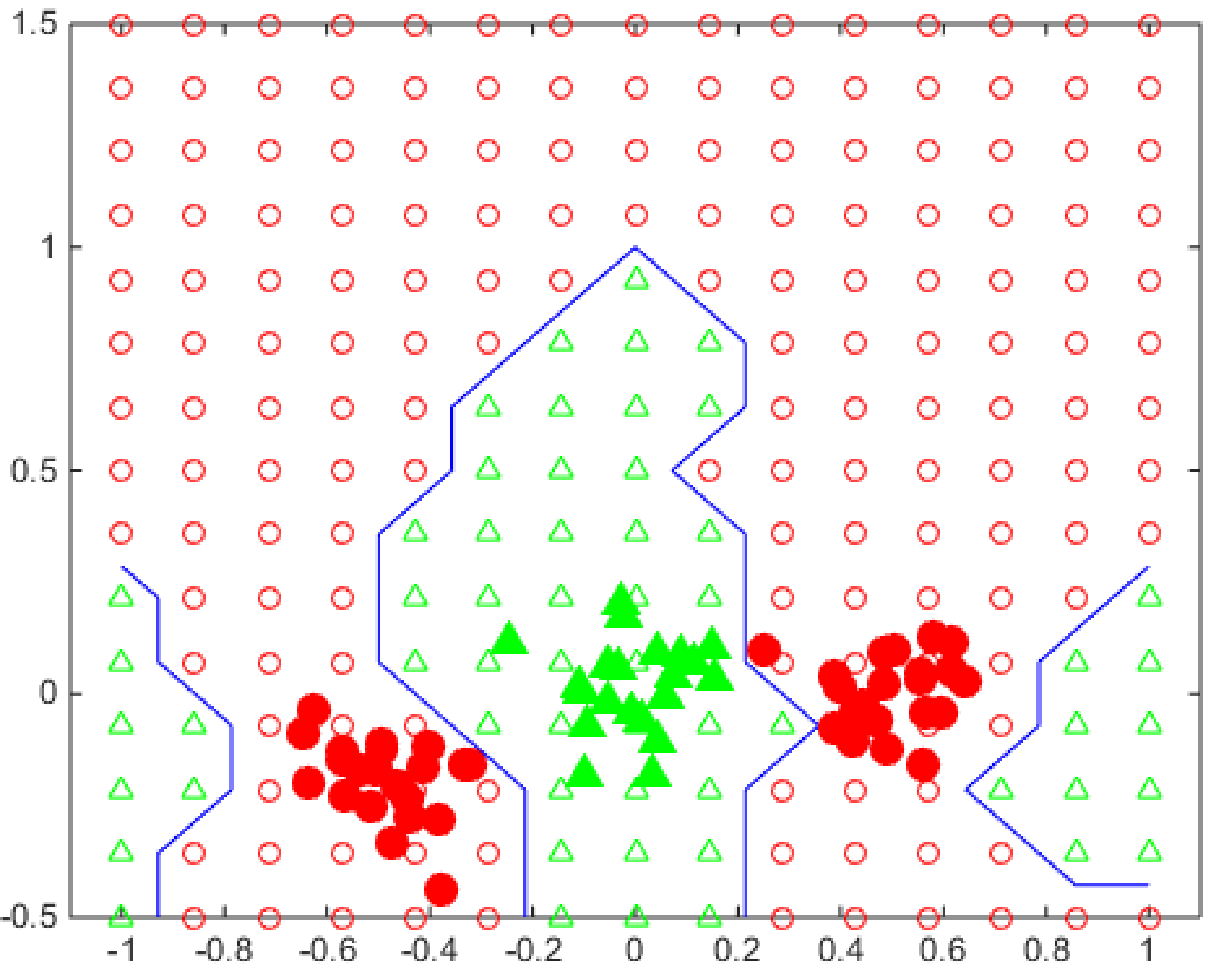}

}
\par\end{centering}

\begin{centering}
\subfloat[]{\includegraphics[scale=0.36]{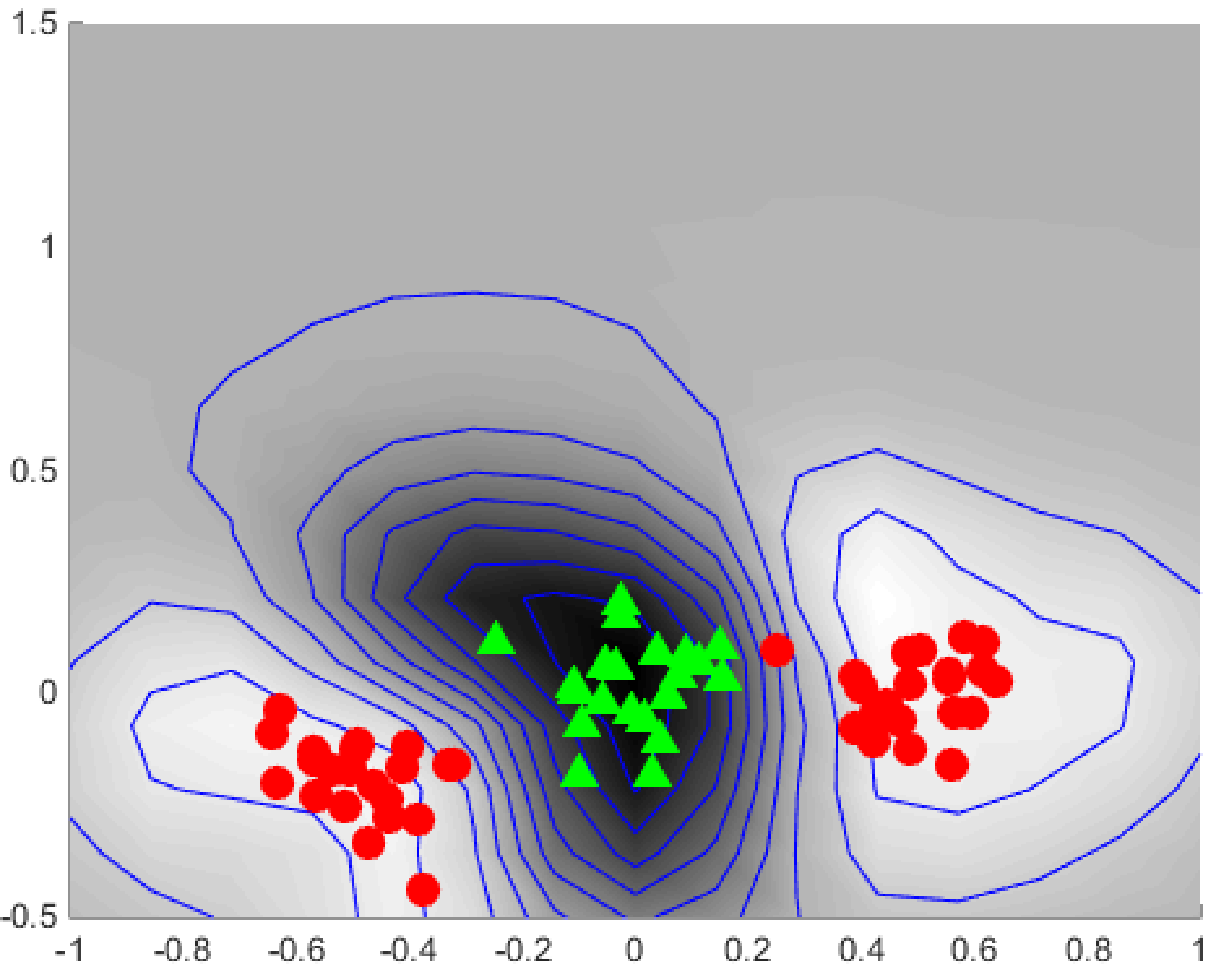}

}\subfloat[]{\includegraphics[scale=0.36]{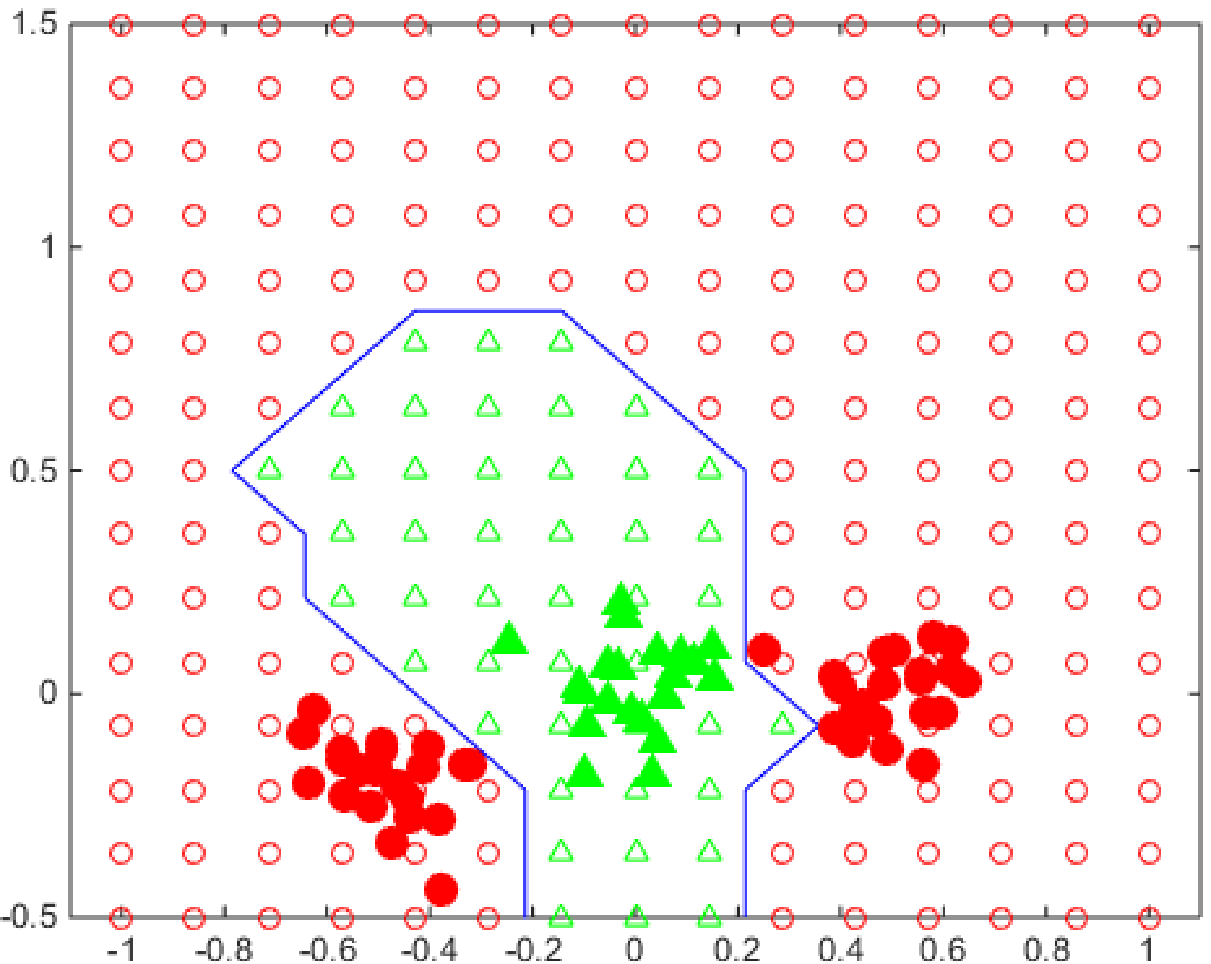}

}
\par\end{centering}

\protect\protect\caption{Classification results for KFDA and Shrinkage KFDA using the kernel
\eqref{eq:kernel} where the number of observations for each Gaussian
$n_{g}$ is equal 20. (a) Contours of constant outputs obtained from
KFDA. (b) Decision boundaries obtained from KFDA. (c) Contours of
constant outputs obtained from shrinkage KFDA. (d) Decision boundaries
obtained from shrinkage KFDA. }
\end{figure}

\section{Conclusions}

In this paper we point out the connections between regularization
of the kernel matrix $\mathbf{K}$ \eqref{eq:Smatrix-1-1} and a shrinkage
estimation of the covariance matrix in the feature space. Since the
kernel matrix $\mathbf{K}$ \eqref{eq:Smatrix-1-1} and the sample
covariance matrix $\mathbf{S}$ \eqref{eq:Smatrix-1} involve inner
and outer products in the feature space, respectively, their spectral
properties are tightly connected. This allows us to examine the kernel
matrix stability through the sample covariance matrix in the feature
space and vice versa. More specifically, the use of kernels often
involves a large number of features when compared to the number of
observations. In this scenario, the sample covariance matrix is not
well-conditioned nor is it necessarily invertible. This forces us
to deal with the problem of estimating high dimensional covariance
matrices under a small sample size. The use of a shrinkage estimator
allows us to effectively address this problem by providing a compromise
between the sample covariance matrix $\mathbf{S}$ \eqref{eq:Smatrix-1}
and the target $\mathbf{T}$ \eqref{eq:Tar1} with the aim of minimizing
the mean-squared error (MSE). Since the spectral properties of $\mathbf{S}$
\eqref{eq:Smatrix-1} and $\mathbf{K}$ \eqref{eq:Smatrix-1-1} are
tightly connected, any modification of the eigenvalues in $\mathbf{S}$
automatically result a related change on the eigenvalues of $\mathbf{K}$
that reflect the optimal correction of the covariance matrix in the
feature space, in the sense of MSE. The result provides an analytical
distribution-free kernel matrix regularization approach that is tuned
directly from the kernel matrix and releases us from dealing with
the feature space explicitly. Numerical simulations demonstrate that
the proposed regularization is significantly effective in classification
tasks.

\appendices{}

\section{Derivation of $\hat{V}(\mathbf{S})$ \eqref{eq:vs-1} and $\hat{V}(\mathbf{T})$
\eqref{eq:vt2} }

The unbiased estimator $\hat{V}(\mathbf{S})$ \eqref{eq:vs-1} is
derived as follow. Let the matrix $\mathbf{S}_{i}$ to be defined
as 
\begin{equation}
\mathbf{S}_{i}=\left(\boldsymbol{\phi}_{i}-\hat{\boldsymbol{\phi}}\right)\left(\boldsymbol{\phi}_{i}-\hat{\boldsymbol{\phi}}\right)^{T},
\end{equation}
where $\hat{\boldsymbol{\phi}}=\frac{1}{n}\sum_{i=1}^{n}\boldsymbol{\phi}_{i}$.
Then $V\left(\mathbf{S}\right)$ can be written as 
\begin{equation}
\begin{array}{c}
V\left(\mathbf{S}\right)=E\left\{ \left\Vert \mathbf{S}-\boldsymbol{\Sigma}\right\Vert _{F}^{2}\right\} =E\left\{ \left\Vert \frac{1}{n-1}\sum_{i=1}^{n}\mathbf{S}_{i}-\boldsymbol{\Sigma}\right\Vert _{F}^{2}\right\} \\
=E\left\{ \left\Vert \frac{1}{n-1}\sum_{i=1}^{n}\left(\mathbf{S}_{i}-\frac{n-1}{n}\boldsymbol{\Sigma}\right)\right\Vert _{F}^{2}\right\} \\
=E\left\{ \frac{1}{\left(n-1\right)^{2}}\sum_{i=1}^{n}\left\Vert \left(\mathbf{S}_{i}-\frac{n-1}{n}\boldsymbol{\Sigma}\right)\right\Vert _{F}^{2}\right\} \\
+E\left\{ \frac{1}{\left(n-1\right)^{2}}\sum_{i=1}^{n}\sum_{j\neq i}^{n}\left\langle \mathbf{S}_{i}-\frac{n-1}{n}\boldsymbol{\Sigma},\mathbf{S}_{j}-\frac{n-1}{n}\boldsymbol{\Sigma}\right\rangle \right\} \\
=E\left\{ \frac{1}{\left(n-1\right)^{2}}\left(1+\frac{1}{n-1}\right)\sum_{i=1}^{n}\left\Vert \left(\mathbf{S}_{i}-\frac{n-1}{n}\boldsymbol{\Sigma}\right)\right\Vert _{F}^{2}\right\} \\
=E\left\{ \frac{n}{\left(n-1\right)^{3}}\sum_{i=1}^{n}\left\Vert \left(\mathbf{S}_{i}-\frac{n-1}{n}\boldsymbol{\Sigma}\right)\right\Vert _{F}^{2}\right\} \\
=E\left\{ \frac{n}{\left(n-1\right)^{3}}\sum_{i=1}^{n}\left\Vert \left(\mathbf{S}_{i}-\frac{n-1}{n}\mathbf{S}+\frac{n-1}{n}\left(\mathbf{S}-\boldsymbol{\Sigma}\right)\right)\right\Vert _{F}^{2}\right\} \\
=E\left\{ \frac{n}{\left(n-1\right)^{3}}\sum_{i=1}^{n}\left\Vert \left(\mathbf{S}_{i}-\frac{n-1}{n}\mathbf{S}\right)\right\Vert _{F}^{2}\right\} +\frac{1}{n-1}E\left\{ \left\Vert \mathbf{S}-\boldsymbol{\Sigma}\right\Vert _{F}^{2}\right\} \\
+E\left\{ 2\frac{n}{\left(n-1\right)^{3}}\frac{n-1}{n}\sum_{i=1}^{n}\left\langle \mathbf{S}_{i}-\frac{n-1}{n}\mathbf{S},\mathbf{S}-\boldsymbol{\Sigma}\right\rangle \right\} ,
\end{array}\label{eq:varSest-2-1-1-1-1}
\end{equation}
where the last term equals zero (observed after entering the sum into
the inner product), which simplifies the expression \eqref{eq:varSest-2-1-1-1-1}
to 
\begin{equation}
V\left(\mathbf{S}\right)=\frac{1}{n-1}V\left(\mathbf{S}\right)+E\left\{ \frac{n}{\left(n-1\right)^{3}}\sum_{i=1}^{n}\left\Vert \left(\mathbf{S}_{i}-\frac{n-1}{n}\mathbf{S}\right)\right\Vert _{F}^{2}\right\} 
\end{equation}
and finally 
\begin{equation}
V\left(\mathbf{S}\right)=E\left\{ \frac{n}{\left(n-1\right)^{2}\left(n-2\right)}\sum_{i=1}^{n}\left\Vert \left(\mathbf{S}_{i}-\frac{n-1}{n}\mathbf{S}\right)\right\Vert _{F}^{2}\right\} .\label{eq:varSest-2-1-1-1-1-1}
\end{equation}

The sum term in $V\left(\mathbf{S}\right)$ \eqref{eq:varSest-2-1-1-1-1-1}
can be modified to 
\begin{equation}
\begin{array}{c}
\sum_{i=1}^{n}\left\Vert \left(\mathbf{S}_{i}-\frac{n-1}{n}\mathbf{S}\right)\right\Vert _{F}^{2}\\
=\sum_{i=1}^{n}\left(\left\Vert \mathbf{S}_{i}\right\Vert _{F}^{2}-2\left\langle \mathbf{S}_{i},\frac{n-1}{n}\mathbf{S}\right\rangle +\left(\frac{n-1}{n}\right)^{2}\left\Vert \mathbf{S}\right\Vert _{F}^{2}\right)\\
=\sum_{i=1}^{n}\left\Vert \boldsymbol{\phi}_{i}-\mathbf{\hat{\boldsymbol{\phi}}}\right\Vert ^{4}-2\frac{n-1}{n}\left\langle \sum_{i=1}^{n}\mathbf{S}_{i},\mathbf{S}\right\rangle +\frac{\left(n-1\right)^{2}}{n}\left\Vert \mathbf{S}\right\Vert _{F}^{2}\\
=\left\Vert diag\left(\mathbf{H}\boldsymbol{\Phi}^{T}\boldsymbol{\Phi}\mathbf{H}\right)\right\Vert _{F}^{2}-\frac{\left(n-1\right)^{2}}{n}\left\Vert \mathbf{S}\right\Vert _{F}^{2}\\
=\left\Vert diag\left(\mathbf{K}\right)\right\Vert _{F}^{2}-\frac{1}{n}\left\Vert \mathbf{K}\right\Vert _{F}^{2}.
\end{array}\label{eq:KK}
\end{equation}
By substituting \eqref{eq:KK} into \eqref{eq:varSest-2-1-1-1-1-1},
the expression in the expectation is therefore the unbiased estimator
$\hat{V}(\mathbf{S})$ \eqref{eq:vs-1}.

In a similar manner to $V\left(\mathbf{S}\right)$ \eqref{eq:varSest-2-1-1-1-1-1},
the expression $V(\mathbf{T})$ can be written as 
\begin{equation}
V(\mathbf{T})=\frac{n}{p\left(n-1\right)^{2}\left(n-2\right)}\sum_{i=1}^{n}E\left\{ \mathrm{Tr}^{2}\left(\mathbf{S}_{i}-\mathbf{S}\right)\right\} .
\end{equation}
Therefore, the unbiased estimator of $V(\mathbf{T})$ is 
\begin{equation}
\hat{V}(\mathbf{T})=\frac{n}{p\left(n-1\right)^{2}\left(n-2\right)}\sum_{i=1}^{n}\mathrm{Tr}^{2}\left(\mathbf{S}_{i}-\mathbf{S}\right),\label{eq:VTT}
\end{equation}
where the sum term in \eqref{eq:VTT} can be written as a function
of the matrix $\mathbf{K}$, i.e., 
\begin{equation}
\sum_{i=1}^{n}\mathrm{Tr}^{2}\left(\mathbf{S}_{i}-\mathbf{S}\right)=\left\Vert diag\left(\mathbf{K}\right)-\frac{1}{n}\mathrm{Tr}\left(\mathbf{K}\right)\mathbf{e}\right\Vert _{F}^{2}.\label{eq:v26}
\end{equation}
Substituting \eqref{eq:v26} into \eqref{eq:VTT} will result in $\hat{V}(\mathbf{T})$
\eqref{eq:vt2}.

\bibliographystyle{IEEEtran}
\bibliography{bib_ICML}

\begin{thebibliography}{10}
\providecommand{\url}[1]{#1}
\csname url@samestyle\endcsname
\providecommand{\newblock}{\relax}
\providecommand{\bibinfo}[2]{#2}
\providecommand{\BIBentrySTDinterwordspacing}{\spaceskip=0pt\relax}
\providecommand{\BIBentryALTinterwordstretchfactor}{4}
\providecommand{\BIBentryALTinterwordspacing}{\spaceskip=\fontdimen2\font plus
\BIBentryALTinterwordstretchfactor\fontdimen3\font minus
  \fontdimen4\font\relax}
\providecommand{\BIBforeignlanguage}[2]{{%
\expandafter\ifx\csname l@#1\endcsname\relax
\typeout{** WARNING: IEEEtran.bst: No hyphenation pattern has been}%
\typeout{** loaded for the language `#1'. Using the pattern for}%
\typeout{** the default language instead.}%
\else
\language=\csname l@#1\endcsname
\fi
#2}}
\providecommand{\BIBdecl}{\relax}
\BIBdecl

\bibitem{scholkopf1998nonlinear}
B.~Sch{\"o}lkopf, A.~Smola, and K.-R. M{\"u}ller, ``Nonlinear component
  analysis as a kernel eigenvalue problem,'' \emph{Neural computation},
  vol.~10, no.~5, pp. 1299--1319, 1998.

\bibitem{Lancewicki2014382}
T.~Lancewicki and M.~Aladjem, ``Locally multidimensional scaling by creating
  neighborhoods in diffusion maps,'' \emph{Neurocomputing}, vol. 139, no.~0,
  pp. 382 -- 396, 2014.

\bibitem{nearestneighbor2014}
L.~Zhang, X.~Zhen, and L.~Shao, ``Learning object-to-class kernels for scene
  classification,'' \emph{Image Processing, IEEE Transactions on}, vol.~23,
  no.~8, pp. 3241--3253, Aug 2014.

\bibitem{scholkopft1999fisher}
B.~Scholkopft and K.-R. Mullert, ``Fisher discriminant analysis with kernels,''
  \emph{Neural networks for signal processing IX}, vol.~1, p.~1, 1999.

\bibitem{scholkopf2002learning}
B.~Sch{\"o}lkopf and A.~J. Smola, \emph{Learning with kernels: Support vector
  machines, regularization, optimization, and beyond}.\hskip 1em plus 0.5em
  minus 0.4em\relax MIT press, 2002.

\bibitem{tikhonov2013numerical}
A.~N. Tikhonov, A.~Goncharsky, V.~Stepanov, and A.~G. Yagola, \emph{Numerical
  methods for the solution of ill-posed problems}.\hskip 1em plus 0.5em minus
  0.4em\relax Springer Science \& Business Media, 2013, vol. 328.

\bibitem{james1961estimation}
W.~James and C.~Stein, ``Estimation with quadratic loss,'' in \emph{Proceedings
  of the fourth Berkeley symposium on mathematical statistics and probability},
  vol.~1, no. 1961, 1961, pp. 361--379.

\bibitem{tradeoff2013}
X.~Lu, M.~Unoki, S.~Matsuda, C.~Hori, and H.~Kashioka, ``Controlling tradeoff
  between approximation accuracy and complexity of a smooth function in a
  reproducing kernel hilbert space for noise reduction,'' \emph{Signal
  Processing, IEEE Transactions on}, vol.~61, no.~3, pp. 601--610, Feb 2013.

\bibitem{bickel}
P.~J. Bickel and E.~Levina, ``\BIBforeignlanguage{English}{Regularized
  estimation of large covariance matrices},''
  \emph{\BIBforeignlanguage{English}{The Annals of Statistics}}, vol.~36,
  no.~1, pp. pp. 199--227, 2008.

\bibitem{rohde2011}
A.~Rohde and A.~B. Tsybakov, ``Estimation of high-dimensional low-rank
  matrices,'' \emph{Ann. Statist.}, vol.~39, no.~2, pp. 887--930, 04 2011.

\bibitem{lancewicki_RL}
T.~Lancewicki and I.~Arel, ``Covariance matrix estimation for reinforcement
  learning,'' in \emph{the 2nd Multidisciplinary Conference on Reinforcement
  Learning and Decision Making}, 2015, pp. 14--18.

\bibitem{lancewicki_control}
S.~Sahyoun, S.~Djouadi, and T.~Lancewicki, ``Nonlinear optimal control for
  reduced order quadratic nonlinear systems,'' in \emph{22nd International
  Symposium on Mathematical Theory of Networks and Systems}, 2016, pp.
  334--339.

\bibitem{lancewicki2015sequential}
T.~Lancewicki, B.~Goodrich, and I.~Arel, ``Sequential covariance-matrix
  estimation with application to mitigating catastrophic forgetting,'' in
  \emph{Machine Learning and Applications (ICMLA), 2015 IEEE 14th International
  Conference on}.\hskip 1em plus 0.5em minus 0.4em\relax IEEE, 2015, pp.
  628--633.

\bibitem{ravikumar2011}
P.~Ravikumar, M.~J. Wainwright, G.~Raskutti, and B.~Yu, ``High-dimensional
  covariance estimation by minimizing l1-penalized log-determinant
  divergence,'' \emph{Electron. J. Statist.}, vol.~5, pp. 935--980, 2011.

\bibitem{icml2014_2}
E.~Yang, A.~Lozano, and P.~Ravikumar, ``Elementary estimators for sparse
  covariance matrices and other structured moments,'' in \emph{Proceedings of
  the 31st International Conference on Machine Learning (ICML-14)}, 2014, pp.
  397--405.

\bibitem{cai2012minimax}
T.~T. Cai and H.~H. Zhou, ``Minimax estimation of large covariance matrices
  under l1 norm,'' \emph{Statist. Sinica}, vol.~22, no.~4, pp. 1319--1349,
  2012.

\bibitem{rajaratnam2008}
B.~Rajaratnam, H.~Massam, and C.~M. Carvalho, ``Flexible covariance estimation
  in graphical gaussian models,'' \emph{Ann. Statist.}, vol.~36, no.~6, pp.
  2818--2849, 12 2008.

\bibitem{khare2011}
K.~Khare and B.~Rajaratnam, ``Wishart distributions for decomposable covariance
  graph models,'' \emph{Ann. Statist.}, vol.~39, no.~1, pp. 514--555, 02 2011.

\bibitem{Fan2008186}
J.~Fan, Y.~Fan, and J.~Lv, ``High dimensional covariance matrix estimation
  using a factor model,'' \emph{Journal of Econometrics}, vol. 147, no.~1, pp.
  186 -- 197, 2008, econometric modelling in finance and risk management: An
  overview.

\bibitem{ledoit2011nonlinear}
O.~Ledoit and M.~Wolf, ``Nonlinear shrinkage estimation of large-dimensional
  covariance matrices,'' \emph{Institute for Empirical Research in Economics
  University of Zurich Working Paper}, no. 515, 2011.

\bibitem{MatrixHandbook}
G.~A.~F. Seber, \emph{A Matrix Handbook for Statisticians}.\hskip 1em plus
  0.5em minus 0.4em\relax John Wiley and Sons, Inc., 2007.

\bibitem{RLDM1}
E.~Theodorou, J.~Buchli, and S.~Schaal, ``A generalized path integral control
  approach to reinforcement learning,'' \emph{J. Mach. Learn. Res.}, vol.~11,
  pp. 3137--3181, Dec. 2010.

\bibitem{slda1}
P.~Bickel, B.~Li, A.~Tsybakov, S.~Geer, B.~Yu, T.~Valdes, C.~Rivero, J.~Fan,
  and A.~Vaart, ``\BIBforeignlanguage{English}{Regularization in statistics},''
  \emph{\BIBforeignlanguage{English}{Test}}, vol.~15, no.~2, pp. 271--344,
  2006.

\bibitem{SMT}
G.~Cao, L.~Bachega, and C.~Bouman, ``The sparse matrix transform for covariance
  estimation and analysis of high dimensional signals,'' \emph{IEEE
  Transactions on Image Processing}, vol.~20, no.~3, pp. 625--640, 2011.

\bibitem{Hoyle}
D.~Hoyle, ``Accuracy of pseudo-inverse covariance learning - a random matrix
  theory analysis,'' \emph{IEEE Transactions on Pattern Analysis and Machine
  Intelligence}, vol.~33, no.~7, pp. 1470--1481, 2011.

\bibitem{Raudys}
S.~Raudys and R.~P. Duin, ``Expected classification error of the fisher linear
  classifier with pseudo-inverse covariance matrix,'' \emph{Pattern Recognition
  Letters}, vol.~19, no. 5 6, pp. 385 -- 392, 1998.

\bibitem{slda2}
L.-F. Chen, H.-Y.~M. Liao, M.-T. Ko, J.-C. Lin, and G.-J. Yu, ``A new lda-based
  face recognition system which can solve the small sample size problem,''
  \emph{Pattern Recognition}, vol.~33, no.~10, pp. 1713 -- 1726, 2000.

\bibitem{kernelQDA}
J.~Wang, K.~Plataniotis, J.~Lu, and A.~Venetsanopoulos, ``Kernel quadratic
  discriminant analysis for small sample size problem,'' \emph{Pattern
  Recognition}, vol.~41, no.~5, pp. 1528 -- 1538, 2008.

\bibitem{optimizationCriterion1}
J.~Ye, R.~Janardan, C.~H. Park, and H.~Park, ``An optimization criterion for
  generalized discriminant analysis on undersampled problems,'' \emph{IEEE
  Transactions on Pattern Analysis and Machine Intelligence}, vol.~26, no.~8,
  pp. 982--994, 2004.

\bibitem{OptimizationCriterion2}
P.~Howland and H.~Park, ``Generalizing discriminant analysis using the
  generalized singular value decomposition,'' \emph{IEEE Transactions on
  Pattern Analysis and Machine Intelligence}, vol.~26, no.~8, pp. 995--1006,
  2004.

\bibitem{slda7}
W.~J. Krzanowski, P.~Jonathan, W.~V. McCarthy, and M.~R. Thomas,
  ``\BIBforeignlanguage{English}{Discriminant analysis with singular covariance
  matrices: Methods and applications to spectroscopic data},''
  \emph{\BIBforeignlanguage{English}{Journal of the Royal Statistical Society.
  Series C (Applied Statistics)}}, vol.~44, no.~1, pp. pp. 101--115, 1995.

\bibitem{slda8}
O.~Ledoit and M.~Wolf, ``A well-conditioned estimator for large-dimensional
  covariance matrices,'' \emph{Journal of Multivariate Analysis}, vol.~88,
  no.~2, pp. 365 -- 411, 2004.

\bibitem{slda12}
J.~Schafer and K.~Strimmer, ``A shrinkage approach to large-scale covariance
  matrix estimation and implications for functional genomics,'' \emph{Statist.
  Appl. Genet. Molec. Biol.}, vol.~4, no.~1, 2005.

\bibitem{Lance}
T.~Lancewicki and M.~Aladjem, ``Multi-target shrinkage estimation for
  covariance matrices,'' \emph{IEEE Transactions on Signal Processing},
  vol.~62, no.~24, pp. 6380--6390, Dec 2014.

\bibitem{lancewicki2014dimensionality}
T.~Lancewicki, ``Dimensionality reduction and shrinkage estimation in
  multivariate data,'' Ph.D. dissertation, BEN-GURION UNIVERSITY OF THE NEGEV
  (ISRAEL), 2014.

\end{thebibliography}

\end{document}